\documentstyle[amstex,epsfig,color,graphicx,titlepage,12pt]{article}

\begin{document}

\renewcommand{\thesection}{\Roman{section}.} \baselineskip=24pt plus 1pt
minus 1pt

\begin{titlepage}
\vspace*{0.5cm}
\begin{center}

\LARGE\bf Comment on \textquoteleft Generalized Heisenberg algebra coherent states for power-law potentials\textquoteright
\\[1.5cm]
\normalsize\bf Shahid Iqbal$^{1,3*}$, and Farhan Saif$^{2,3}$
\end{center}


\vspace{7pt}
\begin{description}


\item  [$^1$]Department of Physics, Govt. College University, Lahore 54000, Pakistan. 
\item  [$^2$]Center for Applied Physics and Mathematics, National University of Science and 
             technology, Islamabad, Pakistan.
\item  [$^3$]Department of Electronics, Quaid-i-Azam University, Islamabad 45320, Pakistan.



\end{description}

\vspace{0.3cm}

\normalsize We argue that the statistical features of generalized coherent states 
for power-law potentials based on Heisenberg algebra, presented in a recent paper 
by Berrada {\it et al}~\cite{gha} are incorrect.

\vspace{0.2cm}

\end{titlepage}

\newpage


In a recent article, Berrada {\it et al}~\cite{gha} constructed
generalized Heisenberg algebra (GHA) coherent states for power-law
potentials \cite{jmp}. Following the earlier analysis of power-law
potentials \cite{sic,sic1}, the authors investigated the GHA coherent states for 
loosely binding potentials and tightly binding potentials.
It is claimed that GHA coherent states for loosely binding
potentials $(k < 2)$ exhibit a super-Poissonian behavior for
lower values of $\vert z\vert$ followed by sub-Poissonian behavior for higher $\vert
z\vert$ values. On the other hand, GHA coherent states of tightly binding potentials $(k >
2)$, always exhibit the sub-Poissonian distribution. However, in this case the states for lower values 
of $k$ (i.e., $k=5$ and $k=10$) are declared less classical as $\vert z\vert$ becomes
large which means that the states in question
get farther from the states exhibiting Poissonian statistics, in
particular Glauber's coherent states \cite{b.gl}. In this comment, we argue that the transition of 
super-Poissonian to sub-Poissonian behavior in GHA coherent states for loosely binding 
potentials and less classical behavior in GHA coherent states of tightly binding potentials for $k=5$ and $k=10$,
presented in ref.~\cite{gha},
are incorrect and are a consequence of wrong numerics. 

In order to discuss the results presented by the authors, we
rewrite the generalized Heisenberg algebra coherent states for
power-law potentials \cite{gha}
\begin{equation}
\left\vert z,k\right\rangle = N(\vert z \vert, k)
\sum_{n=0}^{\infty }\frac{z^{n}}{\sqrt{
g(n,k)}}\vert n\rangle,  \label{gkplp}
\end{equation}
where
\begin{equation}
g(n,k)=\prod_{j=1}^{n}
\left[ \left( j+\frac{\gamma }{4}\right)
^{2k/\left( k+2\right) }-\left( \frac{\gamma }{4}\right) ^{2k/\left(
k+2\right) }\right],  \label{rho}
\end{equation}
and the normalization function
\begin{equation}
N(\vert z \vert, k)=\left( \sum_{n=0}^{\infty }\frac{\vert z\vert^{2n}}{g(n,k)}\right)^{-1/2} .  \label{normk}
\end{equation}
The statistical behavior of GHA coherent states of power-law potentials can be probed
through the weighting distribution, 
\begin{equation}
P_{n}(\vert z\vert,k)\equiv \left\vert \left\langle n\right\vert z,k\rangle \right\vert ^{2}= 
N^{2}(\vert z \vert, k) \frac{\vert z\vert^{2n}}{ g(n,k)}.  \label{w}
\end{equation}
However, in Ref. \cite{gha}, Mandel's $Q-$parameter has been used to determine the nature of weighting distributions of
these states.
This parameter is defined as
\cite{men2}
\begin{equation}
Q=\frac{\sigma^{2}}{\left\langle \hat{N}\right\rangle }-1,
 \label{mq}
\end{equation}
where, $\left\langle \hat{N}\right\rangle$ is the mean and $\sigma^{2}$ is variance, i.e.,
\[\sigma^{2}=\left\langle \hat{N}^{2}\right\rangle-\left\langle \hat{N}\right\rangle^{2}\]
of the corresponding distribution. The Mandel's $Q-$parameter is defined such that, the weighting distribution of
coherent states is Poissonian if $Q=0$, super-Poissonian if $Q>0$ and
sub-Poissonian if $Q<0$. However, it is obvious from Eq.~(\ref{mq}) that
in order to calculate Mandel's $Q-$parameter one needs to calculate the expectation values,
\begin{equation}
\left\langle z, k\right\vert \hat{N}^{m}\left\vert z,k\right\rangle =N^{2}(\vert z \vert, k)\sum_{n=0}^{\infty }n^{m}\frac{\vert z \vert^{2n}}{ g(n,k)
}\equiv\left\langle \hat{N}^{m}\right\rangle,  \label{exp}
\end{equation}
for $m=1,2$, where $\hat{N}$ is bosonic number operator. Therefore, Eq. (\ref{exp}) 
serves as synthesis equation for mean, variance and Mandel's parameter. 

In Figs. 1 and 2 of reference \cite{gha}, Mandel's Q-parameter for GHA coherent states 
of loosely binding potentials $(k<2)$ and tightly binding potentials $(k>2)$, respectively, is plotted 
versus coherent state amplitude $\vert z\vert$. The plots in Fig. 1 show that Mandel's Q-parameter undergoes 
transition from positive values to negative ones after a short range of $\vert z\vert$, afterwards it attains 
a steady state value that approaches to $-1$. As a result, the weighting distribution is reported to undergo a transition 
from super-Poissonian to sub-Poissonian. On the other hand, the plots in Fig. 2 show that Mandel's Q-parameter is 
negative for all values of $\vert z\vert$ (sub-Poissonian distribution), however, for lower values of $k$ 
(i.e., $k=5$ and $k=10$) it sharply tends to reach its minimum value $Q=-1$ at large values of $\vert z\vert$.
The transition of Mandel's parameter from positive values to negative ones in Fig. 1 and a sharp 
deviation from nearly steady state towards its minimum value $Q=-1$ (in case of $k=5$ and $k=10$) in Fig. 2 is
very unexpected. In the following, we present our  analytical as well as numerical arguments against these results.

As a preliminary bit of information, we argue that Mandel's $Q$-parameter depends on 
variance-to-mean ratio ($VMR$) such that $Q=-1$ when $VMR=0$ as  expressed by Eq. (\ref{mq}). In order for 
$VMR=0$ it is required that either variance of
the distribution should be zero ($\sigma^{2}=0$) or mean of the
distribution should take an infinitely large value ($\left\langle
\hat{N}\right\rangle\rightarrow\infty$) in comparison to the variance. 
It is obvious from the synthesis equation (Eq. (\ref{exp})) 
that expectation values of $\hat{N}$ are directly proportional to $\vert z\vert$. Consequently, 
for a finite value of $\vert z\vert$, the mean of the distribution takes finite values and variance takes 
non-zero positive value. Therefore, neither of the
situations that lead to $Q=-1$ is possible in the
case of GHA coherent states for power-law potentials.
In order to validate our analytical argument, we use the synthesis equation (Eq.~(7)) to compute 
the values of mean, variance 
and Mandel's parameter of GHA coherent states of a loosely binding potential defined by $k=1.5$ for various values 
of $\vert z\vert$ which are given in Table 1. We find that, Mandel's $Q$-parameter takes positive values i.e., $0.162$ 
and $0.164$ for $\vert z\vert=12.5$ and $\vert z\vert=12.5$, respectively, in contrast to the results displayed in 
Fig. 1 of Ref. \cite{gha} for $k=1.5$. 
Using the same equation, we can check the values of Mandel's parameter for the plots of other values of $k$ mentioned 
above ($k=0.5, 1$ in Fig. 1 and $k=5, 10$ in Fig. 2). Our analytical and numerical arguments are 
in complete agreement with earlier published results \cite{jmp}. These facts lead us to conclude that the results 
displayed by plots in Fig. 1 and by plots for $k=5, 10$ in Fig. 2 of
\cite{gha} are incorrect.
\begin{table}[ht]
\centering 
\begin{tabular}{c c c c c} 
\hline\hline 
 $\vert z\vert$ & $\left\langle \hat{N}\right\rangle$ & $\sigma^{2}$ & $Q$ & $n_{max}$\\ [0.5ex] 
\hline 
2.5 & 9.44 {\color{red}(9.44)} & 10.05 {\color{red}(10.05)} & 0.064 {\color{red}(0.064)} & 50\\
5.0 & 43.94 {\color{red}(43.94)} & 50.06 {\color{red}(50.06)} & 0.139 {\color{red}(0.139)} & 100\\ 
7.5 & 111.45 {\color{red}(111.43)} & 128.67 {\color{red}(127.79)} & 0.154 {\color{red}(0.147)} & 200\\
10.0 & 216.92 {\color{red}(147.57)} & 251.58 {\color{red}(7.63)} & 0.160 {\color{red}(-0.948)} & 400\\
12.5 & 364.20 {\color{red}(149.14)} & 423.31 {\color{red}(1.56)} & 0.162 {\color{red}(-0.989)} & 600\\
15.0 & 556.57 {\color{red}(149.52)} & 647.64 {\color{red}(0.69)} & 0.164 {\color{red}(-0.995)} & 700\\[1ex] 
\hline 
\end{tabular}
\caption{The values without brackets 
are calculated by taking various values of $n_{max}$, such that, the condition
$n_{max}>n_{th}$ is satisfied, whereas, the values within brackets are calculated for fixed $n_{max}=150$.} 
\label{table:nonlin} 
\end{table}

\begin{figure}
\centering

\includegraphics[width=0.8\textwidth]{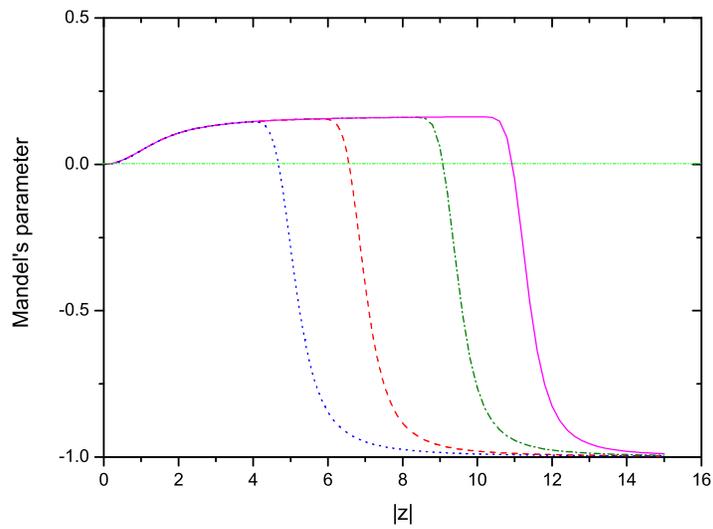}

\caption{Mandel's Q-parameter for $k=1.5$ versus $\vert z\vert$ for $n_{max}=50$ (dotted line),
$n_{max}=100$ (dashed line), $n_{max}=200$ (dashed-dotted line),
$n_{max}=300$ (solid line).} \label{mand}
\end{figure}

As a matter of fact, we probe the possible error in the numerical computation of Mandel's $Q$-parameter 
that led to the wrong conclusions in Ref.\cite{gha}. 
It is seen that synthesis equation (Eq. (\ref{exp})) for mean, variance and Mandel's parameter 
consists of the infinite sum of terms. In numerical computation of mean, variance and Mandel's parameter, 
one has to replace the infinite 
sum involved in Eq. (\ref{exp}) by a 
finite sum. This finite sum is obtained by truncating the summation at a suitable $n=n_{max}$, 
such that, $n_{max} \geq n_{th}$ where 
$n_{th}$ is the threshold beyond which the terms of summation are not contributing 
significantly. For a particular set of coherent state parameters, the threshold value, $n_{th}$, 
for the summation can be probed by hit and trial, such that, the condition $n_{max} \geq n_{th}$ is found to be
satisfied if further increase in the cut off 
value, $n_{max}$, do not change computed result. However, it is worth 
noting that each term of the synthesis equation 
depends on the ratio $\vert z\vert^{2n}/g(n,k)$, therefore, the $n_{th}$ (number of significant terms) of the summation 
increases as $\vert z\vert$ increases. As a result, the 
value of $n_{max}$ should be taken larger while computing $Q$ for greater values of $\vert z\vert$ 
as expressed by Table 1. The situation $n_{max} < n_{th}$, 
where 
summation in the 
synthesis equation is truncated  at a value of 
$n$ beyond which the terms are still contributing significantly leads to the wrong results. The effect of this early 
truncation is indicated 
by Table 1 and Fig.~\ref{mand}.
It is obvious from the Table 1 and the Fig.~\ref{mand} 
that for a particular value of $\vert z\vert$, mean, variance and Mandel's parameter take lower values 
when $n_{max} < n_{th}$ as compared to their corresponding values when all effective terms are included in 
the situation i.e, $n_{max} \geq n_{th}$. 
Moreover, it has been pointed out \cite{jmp} that the dependence of summation 
threshold ($n_{th}$) on $\vert z\vert$ increases sharply as the value of power-law exponent decreases.
Consequently, the computation of Mandel's 
parameter for coherent states of potentials defined by
lower values of power-law exponent 
needs high value of $n_{max}$ to include all effective terms of the summation. As an example, 
we again refer the case of $k=1.5$ given in Table 1, 
where, $n_{th}>650$ for $\vert z\vert=15$ and we have taken $n_{max}=700$ to compute the $Q$, correctly. In order 
to calculate $Q$ for high values
of $\vert z\vert$ (e.g,up to the range in ref. \cite{gha}) a very high value 
of $n_{max}$ would be needed that may exceed the computational limit of ordinary computers. For the reason,
the results presented in Fig. 1 and Fig. 2 ($k=5,10$) for higher values of $\vert z\vert$ are 
questionable as far as their correct computation is concerned. 
However, from Table 1 and Fig. ~\ref{mand} 
it can be inferred easily that if it were possible to perform computation with very high value of
$n_{max}$ so that $n_{max}\geq n_{th}$, then $Q$ would go on increasing without abrupt deviation towards 
$-1$ as $\vert z\vert$ increases. These facts leads us to conclude that the abrupt deviation of Mandel's parameter 
towards negative values displayed by Fig. 1, and the abrupt deviation of Mandel's parameter in plots for
$k=5$ and  $k=10$ towards $-1$ displayed by Fig. 2 is nothing but truncation error, such that, authors truncated the 
infinite sum at $n_{max}<n_{th}$.   

As a conclusion, it is stated that GHA
coherent states for loosely binding potentials $(k < 2)$ exhibit a
super-Poissonian distribution for all values of $\vert
z\vert$ without any transition to sub-Poissonian behavior at higher values of $\vert z\vert$ and GHA
coherent states for tightly binding potentials
$(k > 2)$, exhibit a sub-Poissonian distribution smoothly for all values of $\vert
z\vert$, without showing less classicality (for $k=5$ and $k=10$) at higher values of $\vert z\vert$.
Our arguments presented in this comment are in complete agreement with the earlier published results \cite{jmp}.

\end{document}